# Local Manipulation of Nuclear Spin in a Semiconductor Quantum Well

M. Poggio, G. M. Steeves, R. C. Myers, Y. Kato, A. C. Gossard, and D. D. Awschalom[*]

*Center for Spintronics and Quantum Computing, University of California, Santa Barbara, CA 93106*

**Abstract**

The shaping of nuclear spin polarization profiles and the induction of nuclear resonances are demonstrated within a parabolic quantum well using an externally applied gate voltage. Voltage control of the electron and hole wave functions results in nanometer-scale sheets of polarized nuclei positioned along the growth direction of the well. RF voltages across the gates induce resonant spin transitions of selected isotopes. This depolarizing effect depends strongly on the separation of electrons and holes, suggesting that a highly localized mechanism accounts for the observed behavior.

Nuclear spin has been proposed as a robust medium for quantum information processing[1] in the solid state[2]. Due to the ease with which charge can be controlled in semiconductors, it is natural to use conduction electrons as intermediaries in manipulating nuclear spin. One approach is to tune the population and energy distribution of the electrons[3]; our approach is to directly vary the spatial overlap of spin-polarized electrons with lattice nuclei. The ability to create nanometer-sized nuclear spin distributions combined with long solid-state nuclear spin lifetimes has important implications for the future of dense information storage, both classical and quantum. In addition, control over highly localized interactions between conduction electrons and lattice nuclei may provide a means to manipulate such information.

Here, we use gate voltages to electrically position ~ 8 nm wide distributions of polarized nuclei over a ~ 20 nm range in a single parabolic quantum well (PQW). Optically-injected spin-polarized carriers exploit the contact hyperfine interaction to produce nuclear polarization in the vicinity of their confined wave functions. The thin sheets of polarized nuclei are laterally defined by the diameter of a focused laser spot. Furthermore, the application of resonant RF voltages to the gates provides additional electrical control over nuclear spin. In this case, nuclear depolarization is observed and is attributed to a local charge mediated quadrupolar interaction in contrast to a spin dependent coupling.

The sample[4] is an undoped 100 nm (100) $Al_xGa_{1-x}As$ PQW[5] (Fig. 1a) grown by molecular beam epitaxy. The aluminum concentration $x$ is varied from 7% at the center of the well to 40% in the barriers to create a parabolic potential in the conduction band. Electric fields applied across the gated PQW result, to first order, in the distortion-free displacement of the electron wave function position $z_0$ along the growth direction. Experiments are preformed at 6 K in a magneto-optical cryostat with an applied magnetic field $B_0$ perpendicular to the laser excitation direction. A semi-rigid coaxial cable couples RF voltages to the sample gates.

Time resolved Faraday rotation (TRFR) measurements[6] are preformed using a 76 MHz femtosecond Ti:Sapphire laser tuned near the absorption edge of the PQW (1.62 eV). Laser pulses are split into circularly (linearly) polarized pump (probe) pulses with an average power of 2.5 mW (250 µW). Pulses are modulated by optical choppers at $f_1 = 3.3$ kHz and $f_2 = 1.0$ kHz respectively and are focused to an overlapping spot (~30 µm in diameter) on the semitransparent front-gate. Electron spin precession is well described by

$$\theta_F(\Delta t) = \theta_\perp e^{\frac{-\Delta t}{T_2^*}} \cos(2\pi\nu_L \Delta t + \phi) + \theta_\parallel e^{\frac{-\Delta t}{T_1}}. \qquad (1)$$

where $\theta_\perp$ is proportional to the spin injected perpendicular to the applied field, $\theta_\parallel$ is proportional to the spin injected parallel to the applied field, $T_2^*$ is the inhomogeneous transverse spin lifetime, $T_1$ is the longitudinal spin lifetime. The Larmor frequency $\nu_L = g\mu_B B/h$ depends on the total field $B$ acting on the electrons (i.e. the sum of the applied field and the internal effective nuclear field), the Landé g-factor $g$, the Bohr magneton $\mu_B$, and on Planck's constant $h$.

The Landé g-factor varies with Al concentration $x$, allowing us to track the position of the electron wave function, $z_0$, by measuring $\nu_L$[7]. Fig. 1b shows the dependence of $g$ on external applied gate voltage $U_g$. We use a fit to published experimental data relating $g$ to the Al concentration $x$[8], along with the dependence of $x$ on the growth direction $z$, to plot the dependence of the electron wave function position $z_0$ as a function of $U_g$. For small voltages, the electron and hole form an exciton and the electron wave function position varies little with gate voltage. The data show an electron displacement of 5 nm/V over a ~ 20 nm range (the corresponding calculated hole displacement is -7.5 nm/V). Calculations yield a full-width at half-maximum (FWHM) of the electronic probability distribution, $|\psi(z-z_0)|^2$, of ~ 16 nm.

Spin-polarized photo-excited electrons generate nuclear spin polarization within the PQW through dynamic nuclear polarization (DNP)[9]. DNP, most efficient in semiconductors at liquid helium temperatures[10], acts through the contact hyperfine interaction, written as $A_H\ \mathbf{I} \cdot \mathbf{S} = A_H/2\ (I^+S^- + I^-S^+) + A_H\ I_z\ S_z$, where the hyperfine constant $A_H$ contains the squared modulus of the electron wave function at the position of a nucleus, $\mathbf{I}$ is the nuclear spin, and $\mathbf{S}$ is the electron spin[11]. This 'flip-flop' process is driven by the longitudinal component of electron spin which can be varied by changing the sample angle $\alpha$[12].

The average nuclear polarization $\langle I \rangle$ can be extracted from TRFR measurements of $\nu_L = g\mu_B B_0/h + A_H \langle I \rangle/h$. The measurement of $\nu_L$ and the knowledge of the g-factor and applied field $B_0$, yields the nuclear polarization frequency $\nu_n = A_H \langle I \rangle /h$. For GaAs, calculations show that $\nu_n$ = 32.6 GHz for 100% nuclear polarization[13]; after DNP, $\nu_n$ is measured up to 1 GHz in the PQW corresponding to ~ 2.5% nuclear polarization. Changes in the local nuclear polarization $\langle I \rangle$ within the PQW can be measured directly as changes in precession frequency $\Delta \nu_L$.

To detect nuclear polarization we begin with an unpolarized nuclear lattice; optically pumping the PQW at constant $U_g$ fixes the location of spin-polarized electrons at $z_c$. After 20 minutes TRFR data is taken to determine $\nu_L$ as a function of $U_g$[14]. Comparing Larmor frequencies of the polarized and unpolarized states, we determine $\Delta \nu_L$ vs. $U_g$, shown in Fig 2. The data show localization of the nuclear polarization around the electron wave function's polarizing position $z_c$. Narrow distributions of nuclear polarization (~ 8 nm FWHM) can be created at selected positions within our 100 nm quantum well simply by tuning a DC bias voltage during the polarization process.

A MHz frequency gate voltage causes the periodic displacement of the electron wave function within the PQW introducing of a distribution of frequency components into the electron Larmor precession. The upper curve of Fig. 3a shows spin dynamics described by (1) of an electron at a fixed position $z_0$ (no RF voltage) contrasting the more complex dynamics with an applied RF voltage (lower trace). A simple model is derived to explain the additional frequency components introduced by RF modulation[15]:

$$\theta_F(\Delta t) = \theta_\perp e^{\frac{-\Delta t}{T_2^*}} \cos[\pi(\nu_L(z_{max}) + \nu_L(z_{min}))\Delta t + \phi] J_0[(\nu_L(z_{max}) - \nu_L(z_{min}))\Delta t] + \theta_\parallel e^{\frac{-\Delta t}{T_1}}, \quad (2)$$

where $J_0[x]$ is a Bessel function of the first kind. $\nu_L(z_{max})$ and $\nu_L(z_{min})$ are the Larmor precession frequencies at the maximum and minimum positions sampled by the oscillating electron wave function. Fitting the data with (2) we can determine the maximum ($z_{max}$) and minimum ($z_{min}$) wave function positions as a function of $U_g$ for a range of RF powers (Fig. 3b). Fits to the data are calculated assuming the wave function displacement is governed by the relationship given in Fig. 1b. The only free fitting parameter is the amplitude of RF power across the gates, which, as expected, is found to scale linearly with power applied to the device. The data show that position modulation amplitude, $\Delta z$, varies with $U_g$ at a fixed RF power; $\Delta z$ increases for voltage ranges where the electron moves more easily. This result combined with the excellent agreement of our fitting function (2) with the TRFR data (red line in the lower curve of Fig. 3a) demonstrates our ability to displace the electron wave function over nanometer length scales and on nanosecond time scales.

Applying a resonant RF voltage induces nuclear spin transitions of the polarized nuclei. The induction of these transitions results in sudden drops in the time averaged nuclear spin polarization $\langle I \rangle$ shown in Fig 4b. A decrease in $\langle I \rangle$ leads to a change in $\nu_n$ and thus $\theta_F$ at a fixed delay, whose sign and amplitude depend on our choice of $\Delta t$ and the amount that $\langle I \rangle^{16}$ changes. Nuclear depolarization resonances are apparent for the three most abundant isotopes in the sample, $^{75}$As (7.317 MHz/T), $^{71}$Ga (10.257 MHz/T), and $^{69}$Ga (13.032 MHz/T), at the expected NMR frequencies. The asymmetry of the resonance peaks is due to the long time scales on which DNP acts in this sample; the induction of resonant spin transitions quickly depolarizes ⟨I⟩, however, full re-polarization through DNP takes much longer. Measurements also reveal resonances for each of these isotopes at 1/2, 2/3 and 2 times the nuclear resonance frequencies (not shown). The change in $\theta_F$, approximately proportional to $\Delta v_L$ is strongest for the $2f_{NMR}$ transition, followed in strength by the $f_{NMR}$, $f_{NMR}/2$ and finally the $2f_{NMR}/3$ transition. Resonances at $2f_{NMR}$ indicate the presence of $\Delta m = \pm 2$ transitions in addition to $\Delta m = \pm 1$ transitions (where $m$ is the nuclear spin number along the applied field). The fractional resonances at $1/2f_{NMR}$ and $2/3f_{NMR}$ on the other hand, are a result of these same $\Delta m = \pm 1, \pm 2$ spin transitions induced by harmonics of the RF modulation frequency, which may arise due to nonlinearities in the depolarization mechanism.

Spurious time-varying magnetic fields are ruled out as a depolarizing mechanism due to the low leakage currents between front and back gates as well as a series of control experiments[17]. It is known that $\Delta m = 0, \pm 1, \pm 2$ transitions can occur from interactions of the nuclear quadrupole moment with time-varying applied electric fields modulated on resonance[18]. Though RF voltages across the gates of our sample could induce nuclear quadrupolar resonance (NQR), this effect should persist regardless of the presence of laser-injected carriers in the undoped PQW. In contrast, we find that the application of resonant RF voltage modulation in the absence of laser excitation leads to a greatly reduced $\Delta v_L$ (~ 20% of $\Delta v_L$ at 2.5 mW of average pump power). Additional data (Fig. 4c) show that as laser power increases and more carriers are injected into the PQW, $\Delta v_L$ increases, suggesting more important depolarization mechanisms acting locally in the PQW.

To investigate the spatial extent of the depolarization mechanism, resonant RF oscillations are applied with the electron wave function centered at different positions along $z$. Nuclei are initially polarized at $U_g$ = 0.0 V, then $U_g$ is adjusted to an offset voltage and the electron wave function is oscillated for 20 seconds depolarizing the $^{75}$As nuclei. The RF modulation is then turned off, $U_g$ is restored to its initial value, and $\Delta v_L$ is measured. Fig. 4d shows RF depolarization data where the depolarization amplitude seems to correlate with the displacement of the electron wave function shown in Fig. 1b.

The periodic displacement of the electron probability density under the application of RF voltages varies the local electric and magnetic field landscape acting on the lattice nuclei in the PQW. Resonances due to the modulation of the effective electron magnetic field acting on the nuclei, $\boldsymbol{B}_e = A_H \langle \boldsymbol{S} \rangle / \hbar \gamma_N$, can be ruled out since measurements show that the spin of the optically injected carriers, $\langle \boldsymbol{S} \rangle$, has no effect on the depolarization resonance amplitude. The motion of the electron wave function within the well, however, also produces time-varying electric fields and electric field gradients (EFGs) at the nuclear sites. Calculations of the time-varying fields induced between the electron and hole charge distributions within the well indicate the presence of electric fields and EFGs on the order of $10^6$ V/m and $10^{14}$ V/m$^2$ respectively[19]. Through the quadrupolar moment, these fields will induce $\Delta m$ = 0, ±1, and ±2 transitions at both the fundamental and at twice the NMR frequencies. This local NQR interaction is the most likely candidate responsible for the depolarization resonances observed in our samples.

The experimental data show our ability to control local interactions between electrons and nuclear spin in a PQW with an externally applied gate voltage. Quasistatic bias voltages allow the patterning of nanometer-size nuclear spin distributions and RF voltages periodically displace carriers in the PQW inducing NQR. These depolarization resonances can be controlled both electrically and optically yielding a great degree of flexibility in techniques for coherent nuclear control. The ability to electronically control nuclear spin may be advantageous in quantum information processing[20] and in spintronic devices where nuclei can produce large and localized effective magnetic fields in otherwise non-magnetic materials. We thank R. J. Epstein for helpful discussions and D. C. Driscoll for his MBE expertise and acknowledge support from DARPA, ONR, and NSF.

**Figure Captions**

FIG. 1. (a) Sample orientation with respect to laser excitation and applied magnetic field $B_0$. Sample normal is tilted away from the laser propagation direction by an angle $\alpha = 20°$. A gold pad is annealed to contact the back gate; a semi-transparent layer of gold acts as the front gate. (b) Landé g-factor (left axis) plotted vs. bias voltage $U_g$ in black and central position $z_0$ of the electron wave function (right axis) plotted vs. $U_g$ in red.

FIG. 2. (a) Electron wave function shown schematically, centered at a different polarizing positions $z_c$ (3.27 nm, 7.05 nm, and 10.42 nm). (b) Corresponding nuclear polarization distributions created at $B_0 = 3.98$ T, by polarizing nuclei for 20 minutes at position $z_c$ (blue line). Nuclear polarization is measured as a frequency shift $\Delta v_L$ and is plotted as a function of z (solid points). Red curves are Gaussian fits to the data. Centers of the Gaussian fits are 2.5 nm, 6.3 nm, and 11.0 nm respectively.

FIG. 3. (a) Upper curve: $\theta_F$ as a function of $\Delta t$ with no applied RF voltage (offset 1 mrad for clarity), $B_0 = 6$ T, $U_g = -0.1$ V (fit to equation (1) red). Lower curve: $\theta_F$ as a function of $\Delta t$ at the same $B_0$ and $U_g$ with an off-resonant RF voltage of 0.785 $V_{RMS}$ at 28.5 MHz corresponding to a peak-to-peak oscillation of $z_0$ of ~ 4 nm (fit to equation (2) red). (b) Maximum ($z_{max}$) and minimum ($z_{min}$) wave function positions plotted as a function of $U_g$ for different RMS RF voltages. Square (circular) data points represent the upper (lower) bound of wave function displacement $z_{max}$ ($z_{min}$). Solid lines are fits to $z_{max}$ and $z_{min}$.

Nuclear polarization is constant in (a) and (b); observed effects are explained by electron dynamics alone.

FIG. 4. (a) $\theta_F$ as a function of time delay $\Delta t$. Red "x" indicates $\Delta t = 300$ ps used for scan (b) showing $\theta_F$ as a function of applied gate frequency $f_g$ ((a) and (b): $B_0 = 5.46$ T, RF voltage is 0.14 $V_{RMS}$). Dotted, dashed, long-dashed, and solid vertical lines indicate literature values for $f_{NMR}/2$, $2f_{NMR}/3$, $f_{NMR}$, and $2f_{NMR}$ respectively for each color-coded isotope. Asymmetry in the resonances is due to the slow polarization rate compared to frequency sweep. (c) Larmor frequency shift $\Delta v_L$ for different laser powers during RF irradiation ((c) and (d): $B_0 = 3.98$ T, RF voltage is 0.286, at 29.113MHz for 20 seconds, depolarizing $^{75}$As). (d) $\Delta v_L$ as a function of transient offset voltage where the RF modulation is applied. Bias voltage $U_g$ is always reset to 0.0 V when measuring $\Delta v_L$.

[15] An applied RF gate voltage causes $z_0$, $g$ and the precession frequency $\nu_L$ to rapidly oscillate. Therefore $\cos(2\pi\nu\Delta t + \phi)$ becomes

$$\frac{1}{2\pi}\int_0^{2\pi}\cos\left[2\pi\left(\frac{\nu_L(z_{max})+\nu_L(z_{min})}{2} + \frac{\nu_L(z_{max})-\nu_L(z_{min})}{2}\sin(\tau)\right)\Delta t + \phi\right]d\tau =$$
$$\cos[\pi(\nu_L(z_{max})+\nu_L(z_{min}))\Delta t + \phi]J_0[(\nu_L(z_{max})-\nu_L(z_{min}))\Delta t].$$

Note that the oscillatory motion of the electron wave function is 'stroboscopic' in the sense that the oscillation frequencies, typically between 20 and 100 MHz, are on the order of the laser repetition rate of 76 MHz. Since carriers are present in the undoped PQW for only a few ns after each laser pulse, the motion of the electron wave function is 'strobed' by the pulsed laser.

[16] To detect nuclear resonances, we fix $\Delta t$ (chosen so $\theta_F$ is sensitive to small changes in $\nu_L$) and sweep the applied RF gate voltage frequency $f_g$.

[17] A series of control experiments are carried out in order to ascertain the effect of time-varying magnetic fields created by RF currents near the PQW. A semi-transparent sheet of 50 Å of Ti and 50 Å of Au is evaporated 100 nm above the PQW and RF currents are passed through it. Depolarization resonances are observed at $f_{NMR}$ but not at ½ $f_{NMR}$. Small resonances, likely due to the presence of unintentional time-varying electric fields, are observed at 2 $f_{NMR}$. In order to achieve resonance amplitudes comparable to those observed in this report using these time-varying magnetic fields, currents of about 10 mA (~200 gauss) are required. These currents are orders of magnitude larger than leakage currents (< 100 µA) passing through our sample. It is therefore unlikely that time-varying magnetic fields due to spurious RF currents are responsible for the resonances reported here.

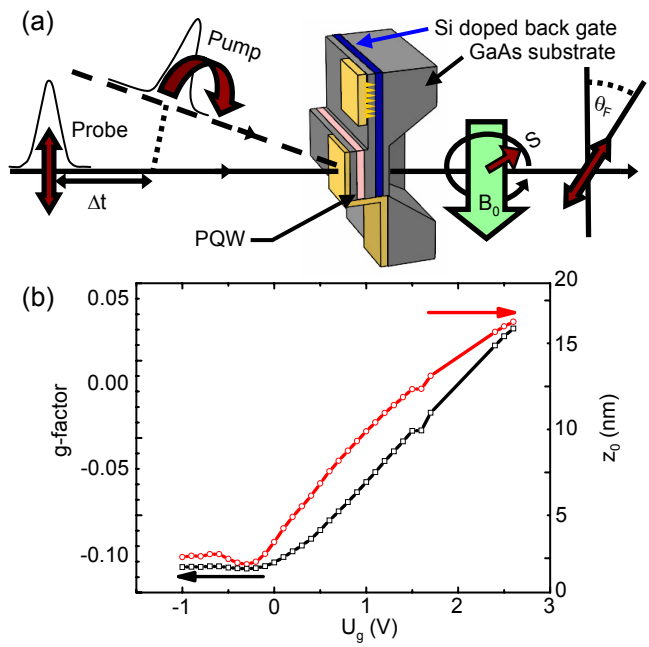

Figure 1. Poggio et al.

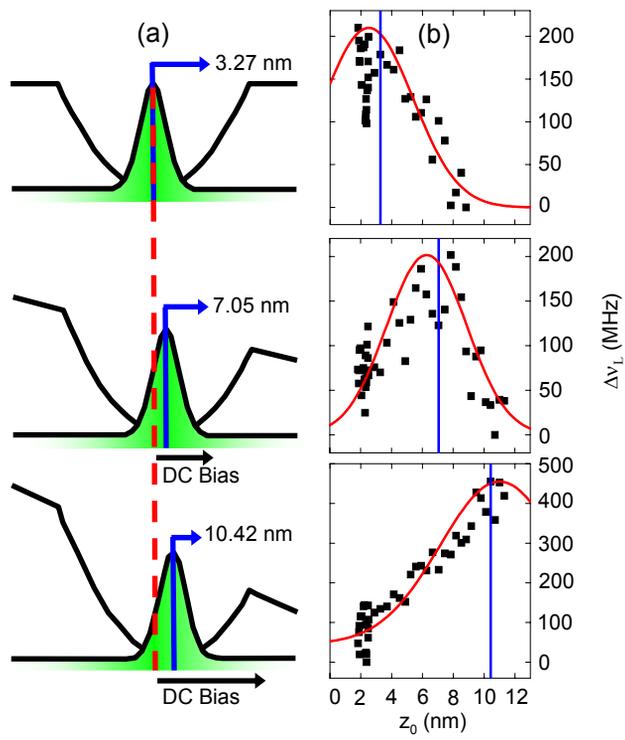

Figure 2. Poggio et al.

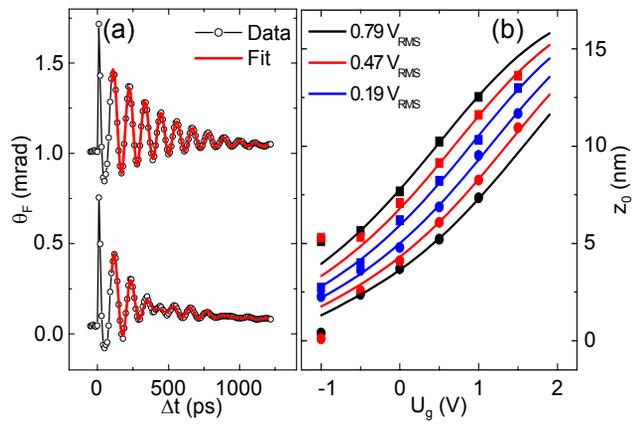

Figure 3. Poggio et al.

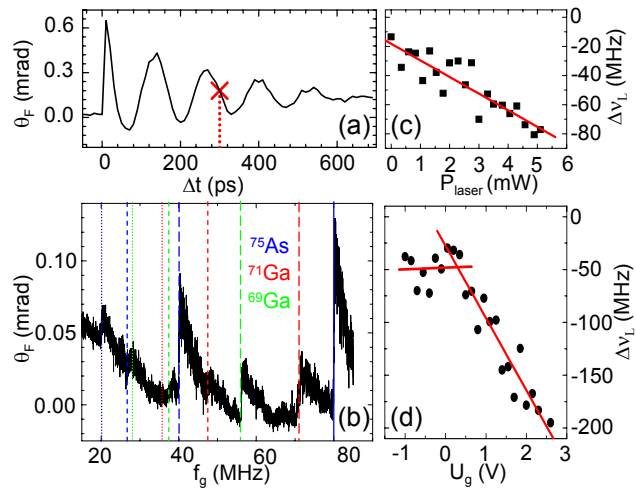

Figure 4. Poggio et al.